\documentclass[]{jfm}
\usepackage{graphicx}
\usepackage{newtxtext}
\usepackage{newtxmath}
\usepackage{natbib}
\usepackage{hyperref}
\hypersetup{
    colorlinks = true,
    urlcolor   = blue,
    citecolor  = black,
}

\newcommand{\RomanNumeralCaps}[1]

\title{Numerical simulation of pulmonary airway reopening by the EOS-based multiphase lattice Boltzmann method}

\author{Bing He\aff{},
  Chunyan Qin\aff{},
  Wenbo Chen\aff{}
 \and Binghai Wen\aff{}
 \corresp{\email{oceanwen@gxnu.edu.cn}}}

\affiliation{\aff{}Guangxi Key Lab of Multi-Source Information Mining \& Security, Guangxi Normal University, Guilin, China, 541004
\aff{}Department of Computer Science and Information Engineering, Guangxi Normal University, Guilin, China, 541004}

\begin{document}
\maketitle

\begin{abstract}
The aerosol formation is associated with the rupture of the liquid plug during the pulmonary airway reopening. The fluid dynamics of this process is difficult to predict because the rupture involved complex liquid-gas transition. Equation of state (EOS) plays a key role in the thermodynamic process of liquid-gas transition. Here, we propose an EOS-based multiphase lattice Boltzmann model, in which the nonideal force is directly evaluated by EOSs. This multiphase model is used to model the pulmonary airway reopening and study aerosol formation during exhalation. The numerical model is first validated with the simulations of \cite{Fujioka2008} and the result is in reasonable agreement with their study. Furthermore, two rupture cases with and without aerosol formation are contrasted and analyzed. It is found that the injury on the epithelium in the case with aerosol formation is essentially the same that of without aerosol formation even while the pressure drop in airway increases by about 67\%. Then extensive simulations are performed to investigate the effects of pressure drop, thickness of liquid plug and film on aerosol size and the mechanical stresses. The results show that aerosol size and the mechanical stresses increase as the pressure drop enlarges and thickness of liquid plug become thicken, while aerosol size and the mechanical stresses decrease as thickness of liquid film is thicken. The present multiphase model can be extended to study the generation and transmission of bioaerosols which can carry the bioparticles of influenza or coronavirus.
\end{abstract}

\begin{keywords}
 bioaerosol, pulmonary airway reopening, equation of state, lattice Boltzmann method
\end{keywords}


\section{Introduction}
\label{sec:intro}
The COVID-19 pandemic caused by the SARS-CoV-2 virus has raged around the world (\cite{Asadi2020}). The fundamental reason for the rapid spread of the coronaviruses is via air-borne aerosols and cause infection through the respiratory tract (\cite{poon2020}). The virus-laden fluid particles (i.e. droplets and aerosols) are expelled from the mouth and nose of an infected person when breathing, talking, coughing, and sneezing (\cite{Mittal2020}). The aerosol emissions into large and small aerosols. Large aerosol particles evaporate more slowly than they settle, contaminating the immediate vicinity of the infected individual. Conversely, small aerosol particles evaporate faster than they settle (\cite{Bourouiba2020}). The aerosol that ranges from sub-micron to hundreds of micrometers is the primary factor in airborne disease transmission. Moreover, the aerosol particle size is closely related to various breathing motions. They are generated in different locations of the pulmonary and carry different concentrations of the virus. Knowledge of the generation mechanisms of aerosol particles and the effect factors that result in the difference of the aerosol particle size constitutes an important basis for exploring the dissemination of infective diseases.

The reopening of closed airways is one of the important mechanisms for aerosol particles generation (\cite{Edwards2004}). The reopening process depends on numerous factors such as  the properties of respiratory tract lining fluid, the geometry of pulmonary airways and the elasticity of the airway wall (\cite{Levy2014}). Much attention has been paid to discuss the reopening process of closed airways and the injury on the airway wall (\cite{Fujioka2008}; \cite{Fujioka2004}; \cite{Fujioka2005}; \cite{Zamankhan2012}; \cite{Vaughan2016}; \cite{Zheng2009}; \cite{Mamba2018}).

\cite{Fujioka2004} and \cite{Fujioka2008} studied the steady and unsteady propagation of the plug in airway and analyze the mechanical stresses on airway walls associated with the plug motion. They further investigated the effects of surfactant (\cite{Fujioka2005}). \cite{Zamankhan2012}focused on the effects of non-Newtonian fluid properties that arise in mucus during airway reopening in pulmonary airways. The splitting of a two-dimensional liquid plug at an airway bifurcation is investigated numerically (\cite{Vaughan2016}). \cite{Zheng2009} analyzed the effects of the wall’s flexibility on the plug propagation in the airway. \cite{Mamba2018} investigated the rupture of a plug driven by a cyclic forcing by  experimental and theoretical method. These previous studies show that numerical modeling the reopening of airways in the lungs still has a fundamental problem. These studies are limited to simulation prior to rupture or just to rupture because they use conventional numerical computation methods which cannot handle complex topological changes during plug rupture.

Some researchers have attempted to develop or apply new interface tracking methods to observe the rupture process and the generation of aerosol particles.  \cite{Haslbeck2010} investigated the aerosol formation by rupture of surfactant films using computations in a fluid dynamics model. They used the high-resolution interface capturing (HRIC) method to track the gas-liquid interface.  \cite{Hassan2011} investigate stresses at the airway walls during plug rupture. They extended an Eulerian-Lagrangian technique to handle topological changes during the plug rupture.  \cite{Malashenko2009} focused on the conditions required for menisci consisting of two gas-liquid interfaces to move unsteadily and disintegrate forming droplets. They employed an algorithm Geometric Reconstruction Scheme (GRS) approach to find air-mucus interface. \cite{Muradoglu2019} investigated the effect of pulmonary surfactants on the propagation and rupture of liquid plugs in the capillary tube by a finite-difference/forward-tracking method. However, these approaches are complicated and time-consuming.

Based on the mesoscopic scale, Lattice Boltzmann model (LBM) can simulate the interaction between different phases and handle the complex interfacial deformation without the interface tracking (\cite{Aidun2010}), LBM-based models are easier to implement than the conventional computational models. A phase-field LBM is been used to simulate the reopening process of closed airways and analyze the effect of capillary number \cite{NingningZhang2016}. In their model, two-phase density ratio is just 10 and the model is two components model. However, the environment of the pulmonary airway is a typical water-vapor system, in which the density ratio is up to 1000. 

Despite tremendous research effort, many aspects of the mechanisms of the airway reopening are still not well understood, especially the key influencers in the size distribution of the aerosols generated are still unclear. In this paper, we investigate the generation of aerosol during the airway reopening by EOS-based multiphase LBM. Our model is a two-phase one-component model and can simulate the large density ratio problem. Thus, the model is closer to the actual situation and the interaction force between the gas-liquid interfaces is directly evaluated by EOSs. The model will be first validated with the simulations of \cite{Fujioka2008}. Furthermore, two rupture cases with and without aerosol formation are contrasted and analyzed. In addition to analyzing the wall stresses during the rupture process, we will additionally study the relation between the various factors and the size distribution of the aerosols generated. 

\section{Method}\label{sec:rules_submission}
\subsection {Lattice Boltzmann method}
Lattice Boltzmann method (LBM) is a mesoscopic numerical simulation method originated from lattice gas automata and kinetic theory (\cite{Chen1998}; \cite{Aidun2010}). It can combine the advantages of many microscopic technologies while still effectively simulating cross-scale dynamics. This feature enables LBM to become an effective method for the numerical simulation of complex fluids. Several collision operators distinguish the variants of the LBE, such as the single-relaxation-time (SRT) model (\cite{qian1992}), the multiple-relaxation-time (MRT) model (\cite{lallemand2000}), the two relaxation-time model (\cite{Ginzburg2008}), and the entropic lattice Boltzmann equation (\cite{Karlin2007}). In particular, the MRT model has great advantages in terms of physical principles, parameter selection, and numerical stability, which has been widely used in many fields (\cite{lallemand2000}). Meanwhile, it has the advantages of LBM, such as simple procedures, high parallel efficiency, and easy handling of complex geometric boundaries. Therefore, it provides an effective solution in dealing with the complex two-phase flow problem during the pulmonary airway reopening of the simulation.

The MRT version can be expressed as
\begin{equation}
{f_i}({\bf{\emph{x}}} + {{\bf{\emph{e}}}_i}{\delta _t},t + {\delta _t}) - {f_i}({\bf{\emph{x}}},t) =  - {M^{ - 1}} \cdot S \cdot {\rm{ }}[{f_i}({\bf{\emph{x}}},t) - {f_i}^{(eq)}({\bf{\emph{x}}},t)].
\end{equation}
\emph{M} is a transformation matrix which linearly transforms the distribution functions to the velocity moments, \emph{S} is a diagonal matrix of nonnegative relaxation times: $S \equiv {\rm{diag(0, - }}{{\rm{\emph{s}}}_e}{\rm{, - }}{{\rm{\emph{s}}}_\varepsilon }{\rm{ ,0, - }}{{\rm{\emph{s}}}_q}{\rm{,0, - }}{{\rm{\emph{s}}}_q},{\rm{ - }}{{\rm{\emph{s}}}_v},{\rm{ - }}{{\rm{\emph{s}}}_v})$. In this paper, the relaxation times are given by \emph{$S{}_e = 1.64$}, $S{}_\varepsilon  = 1.54$, ${S_q} = 1.7$ and $S{}_v = 1/\tau $ (\cite{Mccracken2005}) for the simulations with the MRT LBE. Where  is the particle distribution function at lattice site \emph{x} and time \emph{t}; ${e_i}$ with $i = 0,{\rm{  }}...,N$, \emph{N} is the discrete speed; and $f_i^{(eq)}({\bf{\emph{x}}},t)$ the equilibrium distribution function, where   is the weighting coefﬁcient and u the ﬂuid velocity. 
\begin{equation}
f_i^{(eq)}({\bf{\emph{x}}},t) = \rho{\omega _i}[1 + 3({{\bf{\emph{e}}}_i} \cdot {\bf{\emph{u}}}) + \frac{9}{2}{({{\bf{\emph{e}}}_i} \cdot {\bf{\emph{u}}})^2} - \frac{3}{2}{\emph{u}^2}].
\end{equation}

{\bf EOS-based multiphase flow model} - The phase transition is a more important theme in the multiphase flow system. The phase transition is the macroscopic manifestations of interactions between components or phases on the microscopic scale (\cite{GUO2009}). In the past two decades, many multiphase Lattice Boltzmann models have been proposed. Based on the describing method of interactions between components or phases, multiphase LB models can be divided into the pseudopotential LB method (\cite{SHAN1993}; \cite{Liu2014}), the free energy based LB method (\cite{Swift1995}; \cite{Swift1996}), the phase-field LB method (\cite{He1999}), and the color-gradient LB method (\cite{Gunstensen1991}). Recently, based on free energy theory, \cite{Wen2015} developed a new multiphase LB model in which used pressure tensor to calculate nonideal forces which can accurately describe the interactions between phases. They subsequently improved the model by introducing the chemical potential to calculate nonideal forces and avoiding the calculation of pressure tension divergence (\cite{Wen2017}). Drawing on their methods, based on thermodynamic theory, we proposed a new model. 

In a nonideal fluid systems, the free energy functional within a gradient-squared approximation is (\cite{He2020})
\begin{equation}
\emph{$\Psi$}  =\int{(\psi (\rho ) + \frac{\kappa }{2}|\nabla \rho {|^2})} dx,	
\end{equation}
where $\int {\psi (\rho )} dx$ is the bulk free-energy density at a given temperature with the density $\rho$. And $\int {(\frac{\kappa }{2}|\nabla \rho {|^2})} dx$ gives the free energy contribution from density gradients in an inhomogeneous system, with the surface tension coefficient  $\kappa$. The free energy function in turn determines the diagonal term of the pressure tensor
\begin{equation}
 p(x) = {p_0} - \kappa \rho {\nabla ^2} - \frac{\kappa }{2}|\nabla \rho {|^2},
\end{equation}
where \emph{$p_0$}  is the EOS, and the expression is
\begin{equation}
{p_0} = \rho \psi '(\rho ) - \psi (\rho ),	
\end{equation}
the full pressure tensor can be written as
\begin{equation}
{P_{\alpha \beta }} = p(x){\delta _{\alpha \beta }} + \emph{$\kappa$} (\frac{{\partial \rho }}{{\partial {x_\alpha }}}\frac{{\partial \rho }}{{\partial {x_\beta }}}),
\end{equation}
where \emph{$\delta_{\alpha \beta}$}  is the Kronecker delta function. The excess pressure, namely, the nonideal force, with respect to the ideal-gas expression can be directly computed (\cite{Wen2017})
\begin{equation}
F =  - \nabla  \cdot \mathord{\buildrel{\lower3pt\hbox{$\scriptscriptstyle\leftrightarrow$}} 
\over P} (x) + \nabla  \cdot {\mathord{\buildrel{\lower3pt\hbox{$\scriptscriptstyle\leftrightarrow$}} \over P} _0}(x).
\end{equation}
Where ${\mathord{\buildrel{\lower3pt\hbox{$\scriptscriptstyle\leftrightarrow$}} 
\over P} _0} = \mathord{\buildrel{\lower3pt\hbox{$\scriptscriptstyle\leftrightarrow$}} 
\over I} c_s^2\rho $ is the ideal-gas EOS. From equation (2.4) and (2.6), it can be find
\begin{equation}
{P_{\alpha \beta }} = [{p_0} - \kappa \rho {\nabla ^2}\rho  - \frac{\kappa }{2}|\nabla \rho {|^2}]{\delta _{\alpha \beta }} + \kappa (\frac{{\partial \rho }}{{\partial {x_\alpha }}}\frac{{\partial \rho }}{{\partial {x_\beta }}}).
\end{equation}
From equation (2.8), the divergence of the pressure tensor is written as
\begin{equation}
	\frac{\partial }{{\partial {x_\beta }}}{P_{\alpha \beta }} = [{p_0} - \kappa \rho {\nabla ^2}\rho  - \frac{\kappa }{2}|\nabla \rho {|^2}]\frac{\partial }{{\partial {x_\alpha }}} + \kappa \frac{\partial }{{\partial {x_\alpha }}}(\frac{{\partial \rho }}{{\partial {x_\alpha }}}\frac{{\partial \rho }}{{\partial {x_\beta }}}).	
\end{equation}
After some simple manipulations, can be further written as
\begin{equation}
\frac{\partial }{{\partial {x_\beta }}}{P_{\alpha \beta }} = \frac{\partial }{{\partial {x_\alpha }}}{p_0} - \kappa \frac{\partial }{{\partial {x_\alpha }}}(\rho {\nabla ^2}\rho ) + \kappa \frac{{\partial \rho }}{{\partial {x_\alpha }}}({\nabla ^2}\rho ).
\end{equation}
A relation between the divergence of the pressure tensor and the gradient of the EOS can be obtained by partial integration of equation (2.10),
\begin{equation}
\nabla \mathord{\buildrel{\lower3pt\hbox{$\scriptscriptstyle\leftrightarrow$}} 
\over P} (x){\rm{   \,=\,   }}\nabla {p_0} - \kappa \rho \nabla ({\nabla ^2}\rho ).
\end{equation}
Substituting equation (2.8) into equation (2.11), we propose the EOS-based multiphase flow model by directly evaluate the nonideal force with EOSs.
\begin{equation}
F =  - \nabla {p_0} + \kappa \rho \nabla ({\nabla ^2}\rho ) + \nabla {\mathord{\buildrel{\lower3pt\hbox{$\scriptscriptstyle\leftrightarrow$}} 
\over P} _0}.
\end{equation}

{\bf Proportionality coefficient \emph{k}} - To improve the stability of the multiphase LBM at the large density ratio, a proportional coefficient k was introduced to correlate the dimension unit of the length between the momentum space and mesh space (\cite{Wen2020}),
\begin{equation}
\delta \hat x{\rm{  \,=\,  }}k\delta x.
\end{equation}

Here, the symbols of the quantities with a length dimension are marked with a superscript, including the lattice length, velocity, and nonideal force. The time, density, and temperature are considered to be independent of the length, so they keep the same symbols and values in the two spaces. According to dimensional analyses, this proportional relation uses the following transformations,
\begin{equation}
\hat \psi (\rho ) = {k^2}\psi (\rho ),\hat \psi '(\rho ) = {k^2}\hat \psi (\rho ),{\hat p_0} = {k^2}{p_0},
\end{equation}
\begin{equation}
	\hat \nabla  = {k^{ - 1}}\nabla ,{\hat \nabla ^2} = {k^{ - 2}}{\nabla ^2}	,
\end{equation}
\begin{equation}
\hat F{\rm{ \,=\,  }}kF.
\end{equation}
The LBE of the present model is evolving in the mesh space (the computational mesh).  The EOS and free energy density are calculated in the momentum space, and then they are transformed into the mesh space by the proportional coefficient. Thus, the nonideal force in the mesh space is evaluated by
\begin{equation}
\hat F =  - \hat \nabla {\hat p_0} + \kappa \rho \hat \nabla ({\hat \nabla ^2}\rho ) + \hat \nabla {\mathord{\buildrel{\lower3pt\hbox{$\scriptscriptstyle\leftrightarrow$}} 
\over P} _0}.
\end{equation}

Subsequently, the nonideal force is incorporated into the lattice Boltzmann equation (LBE), which is fully discretized in space, time, and velocity. The nonideal force acts on the collision process by increasing the particle momentum in the equilibrium distribution function, in which the ﬂuid velocity is replaced by the equilibrium velocity ${\bf{\emph{u}}}_{}^{(eq)} = {\bf{\emph{u}}} + \tau F{\rm{/}}\rho $. Correspondingly, the macroscopic ﬂuid velocity is redeﬁned by the averaged momentum before and after the collision ${\bf{\emph{v}}} = {\bf{\emph{u}}} + F{\rm{/2}}\rho $.

{\bf  Verification} - Numerical simulations involving first-order phase transitions and two-dimensional circular droplets have been performed to verify the EOS-based multiphase LBM. The Peng-Robinson (PR) EOS is used to establish a water-vapor system, the computational domain is a 400 × 400 square with periodical boundary condition by using the MRT model. The PR EOS is written as (\cite{Wen2018})
\begin{equation}
{p_0} = \frac{{\rho RT}}{{1 - b\rho }} - \frac{{a\alpha (T){\rho ^2}}}{{1 + 2b\rho  - {b^2}{\rho ^2}}}.	
\end{equation}
Where the temperature function is $\alpha (T) = {[1 + (0.37464 + 1.5422{\omega ^2}) \times (1 - \sqrt {T/{T_c}} )]^2}$, and the acentric factor \emph{$\omega$}  for water and methane are 0.344 and 0.011, respectively. The universal gas constant is \emph{R} = 1, the attraction parameter is \emph{a} = 2/49 and volume correction is \emph{b} = 2/21. The relaxation time is 1.2 and the proportional coefficient is \emph{k} = 0.1. The parameter \emph{$\kappa$} = 0.001 in the equation(2.17) for the model, unless otherwise specified. The numerical implementation of the present model is discussed next.

It is significant to evaluate the thermodynamic consistency of the present model by comparing the two-phase coexistence curve obtained from the simulations with the theoretical curve predicted by the Maxwell equal-area construction. The middle part of the computational domain is initialized as a circular droplet with radius \emph{R} = 30 lattice units, while the remaining part is gas. PR (water) EOS and PR (methane) EOS are used in our simulations, each of which evolved in 150,000 steps. The simulation results are shown in figure 1(\emph{a}) and (\emph{b}). When the temperature \emph{Tr} = 0.3, the calculation results of the present model are still consistent with the Maxwell equal-area construction, indicating that the present model still has thermodynamic consistency at low temperatures.
\begin{figure} 
  \centerline{\includegraphics{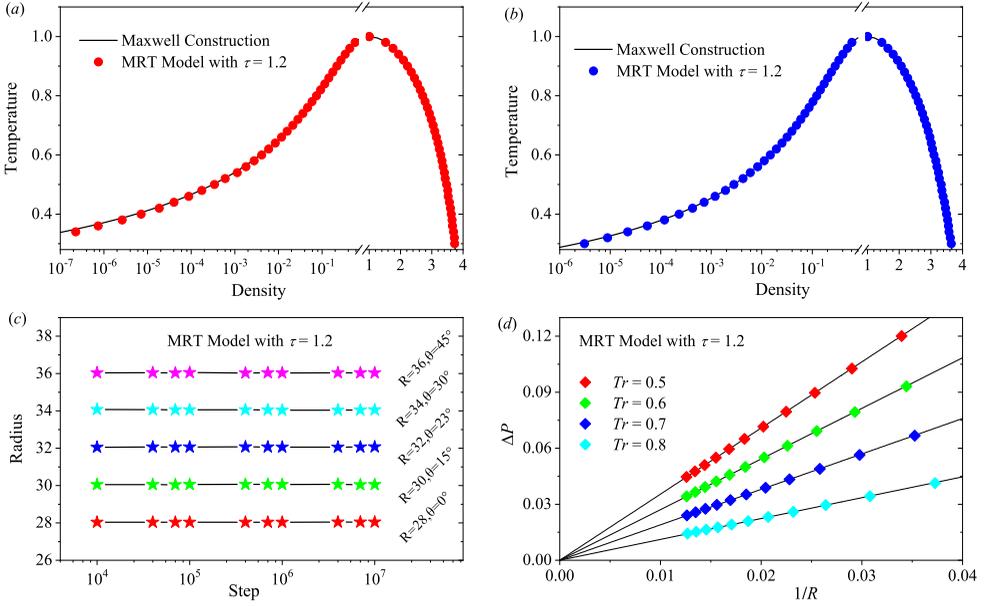}} 
  \caption{(\emph{a}) Gas-liquid density coexistence curve of PR (water) EOS; (\emph{b}) Two-phase density coexistence curve of PR (methane) EOS; (\emph{c}) Test of Galilean invariance in multi directions; (\emph{d}) Test of the Laplace law at multiple temperatures, the solid line has a slope equal to the surface tension. }
\label{fig:ka}
\end{figure}

We simulate circular droplet at \emph{Tr} = 0.6 and in different reference frames and verify the Galilean invariance of the current model by using the degree of surface circle. Taking the symmetry of LBM into consideration, the verification velocity direction is 0°, 10°, 23°, 30°, and 45°, while the corresponding droplet radius is 28, 30, 32, 34, and 36 lattice units, at the same time the reference frame velocity is 0.02 lattice units at each moment. As shown in figure 1(\emph{c}), the radius of the droplet remains the same when it evolves from 10,000 steps to 10,000,000 steps. Numerical results show that the current model has Galilean invariance.

Surface tension is the basic physical property of gas-liquid interface. The method of \cite{Li2021} is employed to adjust the surface tension. The Young-Laplace equation is generally used to evaluate the influence of surface tension in multiphase flows, and it is also an important condition to verity the feasibility of the multiphase flow model. The Young-Laplace equation shows that when the droplet is in a steady state with the surrounding gas, the pressure difference $\Delta$\emph{P} between inside and outside the droplet and the radius \emph{R} satisfies,\begin{equation}
\Delta P = {P_{in}} - {P_{out}} = \sigma /R,	
\end{equation}
where $\sigma$ represents the surface tension of the droplet, \emph{$ P_{in} $} and \emph{$ P_{out} $} are the internal pressure and external pressure of the droplet, respectively. The droplet was placed in the middle region of the flow field. Following the figure 1(\emph{d}), the droplet radius is changed from 25 to 80 lattice units at 5 lattice units intervals. The numbers of evolution of each experiment are 100,000, the temperature change from 0.5 to 0.8 at 0.1 intervals. The simulate results show that the pressure difference between the internal and external of the droplet increases linearly with the reciprocal of the droplet radius increases, and each black line has a stable slope. Therefore, the experimental verification that the present model satisfies the Young-Laplace equation.

\subsection {Pulmonary airway model}
Based on PR EOS of the water, MRT-LBM is used for simulate the reopening of pulmonary airway. A schematic of closed pulmonary airway is shown in figure 2. The pulmonary airway is simplified to a circular rigid tube of radius \emph{R}. The liquid film is adsorbed to the airway walls, the liquid plug is located in the middle of airway. The blue region represents the liquid, while the white region represents the air. The red line represents the trend of air-liquid interface under the pressure. The computational domain is a rectangular with a length of 1500 and a width of 200 lattice units. The pulmonary airway radius \emph{R} is 100 lattice units, which corresponds to 300 $\mu$m at the macroscopic scale and $ 13^{th} $ generation bronchioles (based on the human pulmonary airway model (\cite{Carrington1965}). The liquid film thickness is denoted as \emph{h}. The liquid plug thickness is denoted as \emph{b}. Extrapolated condition is applied on the left and right sides, while the no-slip boundary condition is adopted on the top and bottom boundaries. The relaxation time is \emph{$ \tau $} = 1.2, and the temperature is \emph{Tr} = 0.6, in which case the density ratio of two-phases nears to 1000. The corresponding density of air phases is \emph{$ \rho_{g} $} = 1 g/$ cm^3$, while the density of liquid phases is \emph{$ \rho_l $} = 1000 g/$ cm^3$. The viscosity is $ \mu  $  = 6.947$\times$ $ 10^{-4} $ Pa·s. A constant surface tension is assumed. The pressure drop between the front and rear air phases induced by inhaled air is important factor during the airway reopening process. To mimic the pressure drop, the body force which is denoted as \emph{F} is applied to the flow field in the airway. The body force, \emph{F}, drives the liquid plug movement.
\begin{figure}
  \centerline{\includegraphics{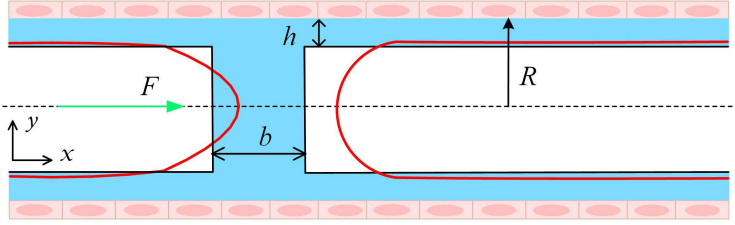}} 
  \caption{Schematic presentation of the closed airway}
\label{fig:ka}
\end{figure}
\begin{figure}
  \centerline{\includegraphics{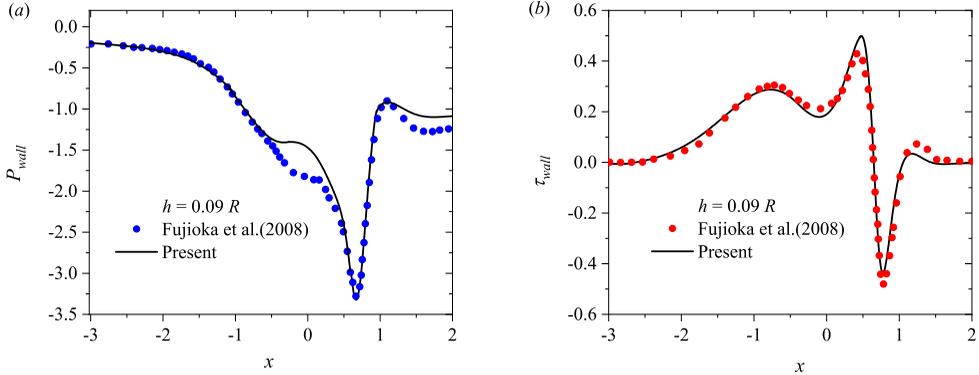}} 
  \caption{Wall pressure and shear stress distribution}
\label{fig:ka}
\end{figure}

To compare with the previous study of \cite{Fujioka2008}, which focused on the plug dynamics prior to the rupture, the same initial condition is employed. The initial liquid plug thickness is \emph{b} = 1 \emph{R},the front liquid film thickness is 0.1 \emph{R} while the tail liquid film thickness is $\emph{h} \approx 0.09 $ \emph{R}. The force is \emph{F} = 0.3 $\mu$dyn. The liquid plug is bent out of shape under the force and the thickness decreases. When the liquid plug thickness \emph{b} decreased to 0.3 \emph{R}, the wall pressure and the wall shear stress distributions are showed as figure 3. The horizontal axis is a relative coordinate which is relative to the middle of the liquid plug as \cite{Fujioka2008}. The wall pressure and wall shear stresses have both negative and positive peaks around the front meniscus region. Result is in reasonable agreement with the study of \cite{Fujioka2008}.

\section {Results and Discussion}
Under various kinematical conditions that correspond to different expiratory activities with various pressure drops, a liquid plug in small lung airways may rupture in different form. It may rupture forming small aerosol droplets, or not. The two cases are discussed respectively. 

\subsection {Rupture without aerosol formation}
\begin{figure}
  \centerline{\includegraphics{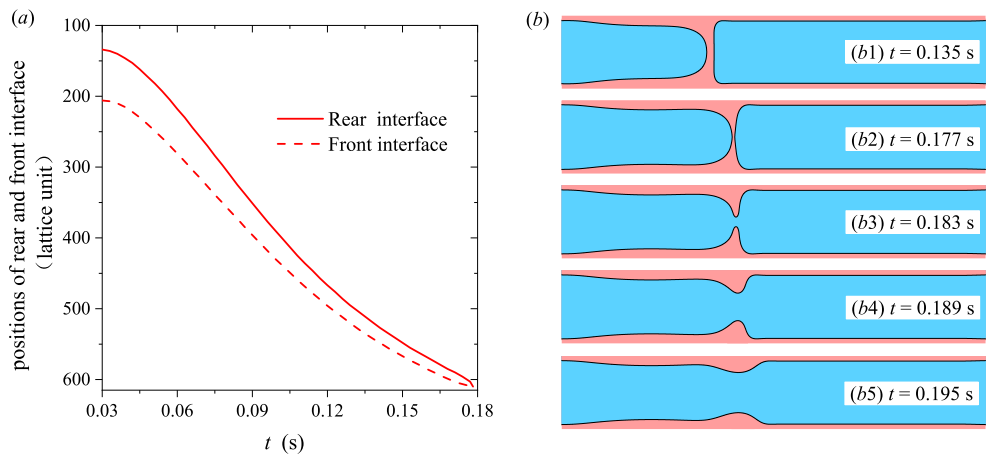}} 
  \caption{Pulmonary airways reopen without aerosol formation (\emph{a}) positions of front and rear interface of liquid plug (\emph{b}) rupture process of a liquid plug along the airway}
\label{fig:ka}
\end{figure}

The thickness of liquid film is commonly slightly thick in unhealthy lungs due to produce excessive bronchial secretions. In order to observe the unhealthy lungs, the thickness considered in the present study is 0.15 of the airway radius, 0.15 \emph{R}. The initial liquid plug which remains same is equal to the airway radius, 1 \emph{R}. The body force is 0.9 $\mu$dyn. The airway reopening includes two stages with different dominant force: prior to rupture and after rupture. At the initial stages, the pressure drop is the primary force driving the liquid plug. The liquid plug begins to deform and move forward under the force of pressure. The positions of front and rear interface of liquid plug are shown in figure 4(\emph{a}). It is obviously that the rear interface moves faster than the front interface. The cross‐sectional shape of liquid plug changes from rectangular to hyperboloidal. The thickness of liquid plug is gradually getting thinner during the interface motion. Figure 4(\emph{b}) shows the rupture process in detail. Prior to rupture, it can be observed that the rear air finger is more sharply than the front air finger and the thickness of the rear liquid film increases (figure 4(\emph{b}1)). When the liquid plug moves forward, the plug ceaselessly leaves liquid to the rear liquid film. As this process proceeds it would result in the thickness of liquid plug decreases to a critical value. Under the body force and surface tension, it ultimately ruptures in the middle at near 0.177 s. After the rupture, surface tension is the primary force driving the interface retraction. The tips of the up and down broken interfaces become round and shrink. About 0.02 s, the up and down broken interfaces retract to the liquid layer lining the airway walls (figure 4 (\emph{b}3 $\sim$ \emph{b}5)). 
\begin{figure}
  \centerline{\includegraphics{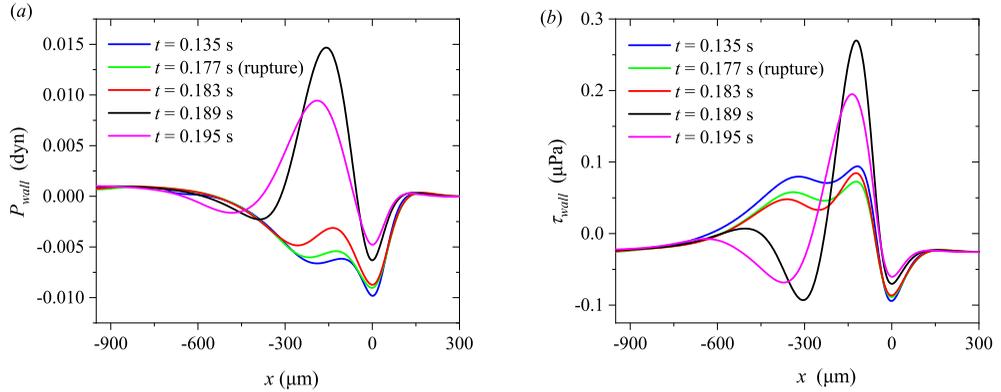}} 
  \caption{Variations of the wall shear stresses at the airway walls during liquid plug rupture process without aerosol formation}
\label{fig:ka}
\end{figure}
\begin{figure}
  \centerline{\includegraphics{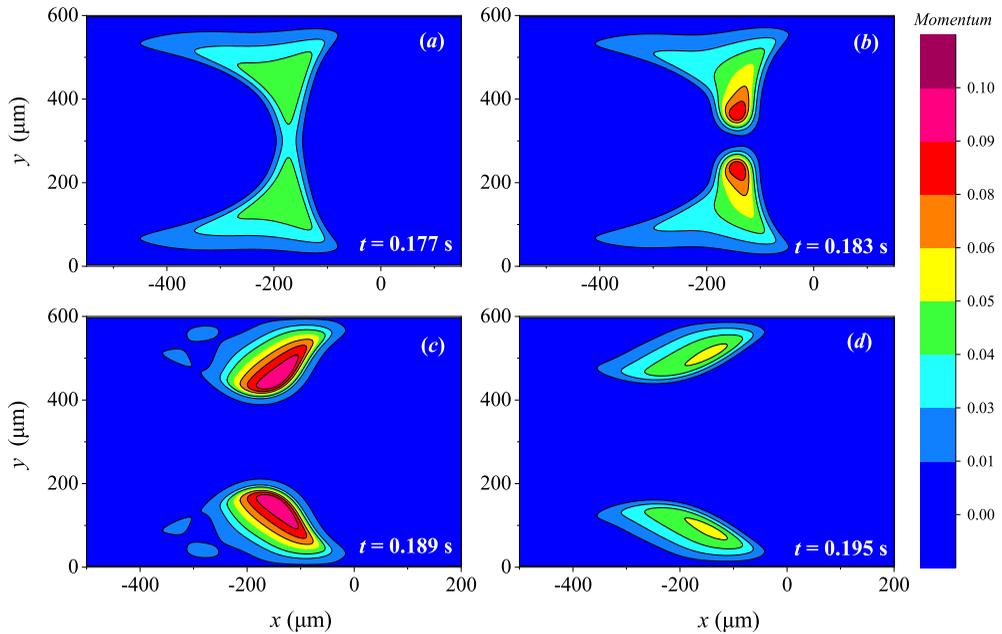}} 
  \caption{Snapshots of momentum distributions during liquid plug rupture without aerosol formation (\emph{a}) \emph{t} = 0.177 s (\emph{b}) \emph{t} = 0.183 s (\emph{c}) \emph{t} = 189 s (\emph{d}) \emph{t} = 0.195 s}
\label{fig:ka}
\end{figure}

During the reopening of airway, the pressure and wall shear stress on the region of the liquid plug appear large variations. These variations are shown in figure 5 in where the origin fixed at the location of minimum liquid film thickness in order to compare with the case with aerosol formation. It can be observed that the pressures and the wall shear stresses have both negative and positive peaks around the liquid plug region. The negative peaks always occur at the location of minimum liquid film thickness. Before the rupture, that is $ t \leq 0.177 $ s, the negative peak values of pressure are close to -0.009 dyn, while that of wall shear stresses are -0.089 $\mu$Pa. After the rupture, the negative peak values decrease in magnitude little by little, while the positive peaks increase in magnitude. The maximum positive peaks reaching \emph{$P_{wall}$} = 0.0147 dyn and \emph{$\tau_{wall}$} = 0.269 $\mu$Pa occur at t = 0.189 s. The different between the maximum and minimum of the pressure is 0.021 dyn, while that different of the wall shear stress is 0.36 $\mu$Pa. The location in which positive peak occurs lags behind the location of minimum liquid film thickness. After a period of time, negative and positive peaks disappear with the disappearing of the ruptured plug. The results are in reasonable agreement with the study of \cite{Hassan2011};  \cite{Malashenko2009}; \cite{Muradoglu2019}. The results also explain why the epithelial cell suffered injury during the airway reopening process.

There has been a scientific consensus about that the variation of the wall shear stress leads to a detrimental effect on the cells lining the airways. But the reason why the maximum positive peak occurs after the rupture isn’t concerned. In order to find out the reason, the momentums of liquid plug region at various times are calculated. The snapshots of momentum are illustrated in figure 6. It is can be found that the broken plug retracts quickly to the wall liquid film in about 0.02 seconds. Before the rupture, the momentum is accumulated in deformed liquid plug as shown in figure 6(\emph{a}). After the rupture, the largest momentum collects in the tip of broken liquid plug under surface tension as shown in figure 6(\emph{b}) and (\emph{c}). The large momentum reaches the airway wall following the tip of broken liquid plug retracts. It indicated that the maximum positive peaks of the pressure and wall shear stress mainly is resulted by the impact of the broken liquid plug. 
 
\subsection {Rupture with aerosol formation}
A vast volume of literature has discussed the liquid plug rupture without aerosol formation under various conditions. There is a paucity of literature on the liquid plug rupture with aerosol. \cite{Malashenko2009} find that the liquid plug can break up to form droplets during the airway reopening process. But they don’t analyze the variations of force on the wall during this process. In here, the case with aerosol formation is considered. The initial liquid plug and liquid film thickness remain same as in the previous section, the body force increases to 1.5 $\mu$dyn. The airway reopening process is shown in figure 7. This process also contains two stages. Before the rupture, the pressure drop similarly makes the liquid plug deform and move forward. Similarly, the rear interface moves faster than the front interface as shown in figure 7(\emph{a}). And the thickness of liquid plug is getting thinner. But the shape of liquid plug under greater pressure is different from the case that is rupture without aerosol. The shape is no longer hyperboloidal, but rather meniscus shape as shown in figure 7(\emph{b}1). The increased pressure causes that the plug fast moves and the curvatures of upper and lower ends in the front interface become larger. Due to lacking liquid supplement, two necks occur in the liquid plug as shown in figure 7(\emph{b}2). It is obvious that the rupture occurs more earlier than the case without aerosol formation. Two necks ultimately break after 0.12 s and a non-spherical droplet generates. After the rupture, the droplet shrinks into a spherical droplet and two remainders of liquid plug also shrink into the wall under the surface tension (figure 7(\emph{b}3 $\sim$ \emph{b}5)).
 \begin{figure}
  \centerline{\includegraphics{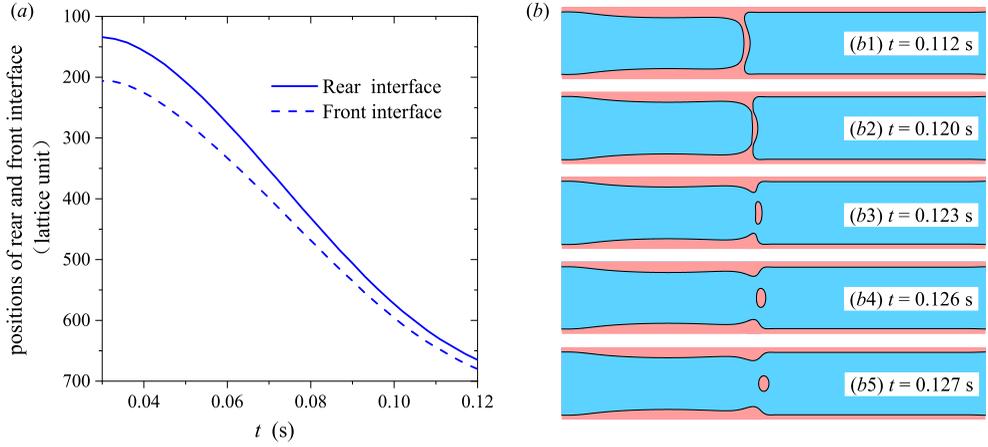}} 
  \caption{Pulmonary airways reopen with aerosol formation (\emph{a}) positions of front and rear interface of liquid plug (\emph{b}) rupture process of a liquid plug along the airway}
\label{fig:ka}
\end{figure}
 \begin{figure}
  \centerline{\includegraphics{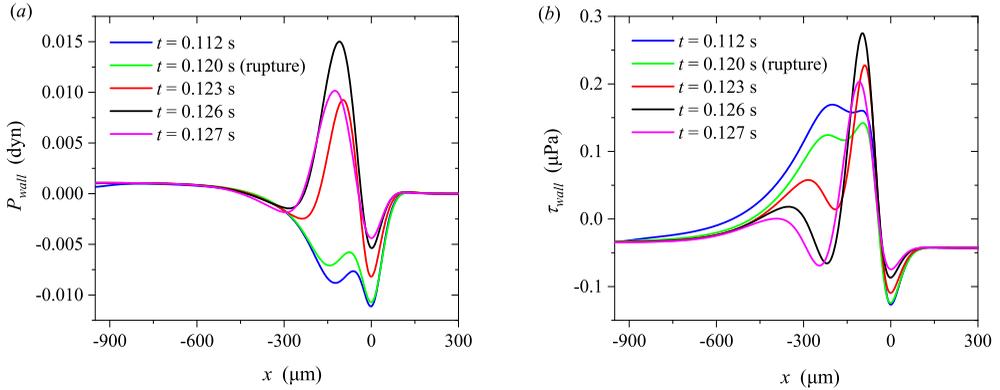}} 
  \caption{Variations of the pressure and the wall shear stresses at the airway walls during liquid plug rupture process in the case with aerosol formation.}
\label{fig:ka}
\end{figure}

During the process of airway reopen in this case, the airway wall also undergoes great the pressure and the wall shear stress. In figure 8, similar variations of the pressure and wall shear stress at the airway walls are observed as liquid plug rupture process in that case without aerosol formation. It is different that the negative peak in magnitude is slightly larger than that of the previous case. In this case the curvature of the front interface of liquid plug is larger than that of the previous case before the rupture, and hence the negative peak becomes large. However, the body force \emph{F} enlarges by about 67\%, the difference between maximum and minimum values of wall pressure and wall shear stress increase only about 6.7\% and 10\%, respectively. This means that the flow-induced injury on the epithelium do not obviously increase in the case with aerosol formation, even while the pressure drop in airway obviously enlarges. 
\begin{figure}
  \centerline{\includegraphics{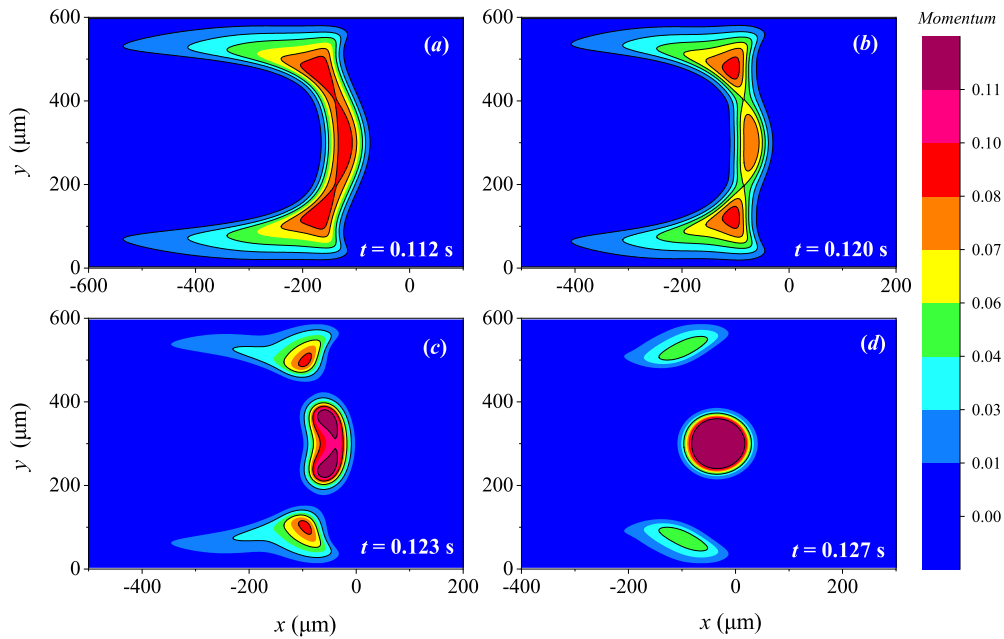}} 
  \caption{Snapshots of momentum distributions during liquid plug rupture with aerosol formation (\emph{a}) \emph{t} = 0.112 s (\emph{b}) \emph{t} = 0.120 s (\emph{c}) \emph{t} = 0.123 s (\emph{d}) \emph{t} = 0.127 s.}
\label{fig:ka}
\end{figure}

The momentum distributions during liquid plug rupture with aerosol formation are shown as figure 9. It can be observed that the deformed liquid plug has accumulated a large momentum before the rupture. The momentum is divided into three parts following two necks occur. After the rupture, the middle part that forms the droplet takes most of the momentum away. The remaining two parts of the liquid plug retract to the airway wall liquid film, like the case without aerosol formation. Due to the momentum on the remaining two parts are obviously diminish, the impact of the broken liquid plug is weaken. That is, the flow-induced injury on the epithelium reduces. This can be used to explain why the difference between maximum pressure and minimum values of the wall pressure and wall shear stress in the case without aerosol formation is a little larger than that of the case with aerosol formation (See in figures 6 and 8), even under the more large pressure drop. The result is agreement with the experimental result in \cite{Malashenko2009}.

\subsection {Factors affecting aerosol size}
The aerosol size depends on respiratory activity and shows a high individual variability (\cite{Almstrand2010}). The pressure drop in airway influences droplet size, the thickness of liquid plug and liquid film also influence droplet size. Further simulations are performed to observe these effects.

The pressure drop in airway may vary with age or due to illness (\cite{Magniez2016}). Its effect is considered within a wide range \emph{F} $ 2 \sim  5$ $ \mu $dyn, corresponding to the pressure drop $1 \sim 15$ cm$H^2O$ which are known to exist in the lung. The results are shown in figure 10. Two group simulations in which the thickness of liquid plug is 1 \emph{R} and 0.6 \emph{R} respectively are carry out. The thickness of liquid film all is 0.15 \emph{R}. Figure 10(\emph{a}) shows that the aerosol mass is a distinct increase as increase of the body force. The results also show that the thickness of liquid plug also has a significant influence on the aerosol mass under the same body force. Furthermore, the effects of pressure drop on the different between the maximum and the minimum of \emph{$P_{wall}$} and \emph{$\tau_{wall}$} are analyzed as shown in figure 10(\emph{b}). The different values of \emph{$P_{wall}$} increase more than double and the different values of \emph{$\tau_{wall}$} also increase by about 80\% as pressure drop in airway changes from 2.0 $ \mu $dyn to 5.0 $ \mu $dyn. It is obvious that the larger pressure drop is more prone to the cell injury. 
\begin{figure}
  \centerline{\includegraphics{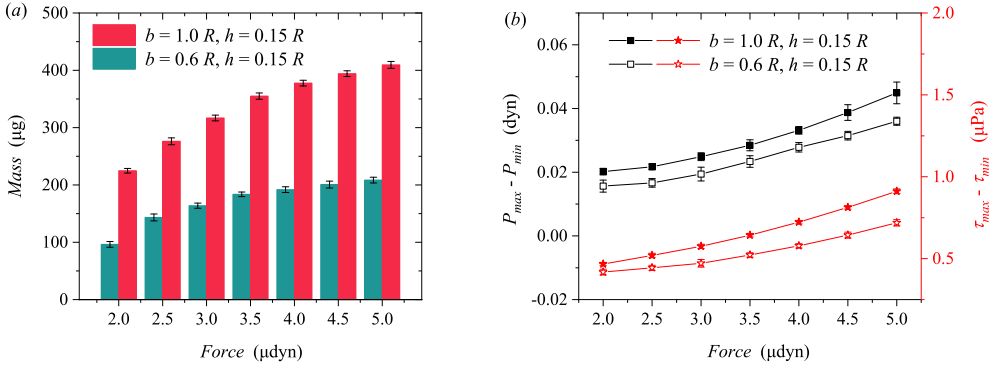}} 
  \caption{Effects of thickness of liquid plug \emph{b} on aerosol size (\emph{a}) the relationship between \emph{b} and aerosol quality (Effects of different pressure drop on aerosol size (\emph{a}) the relationship between pulmonary airway pressure and aerosol quality (\emph{h} = 0.15 \emph{R}, \emph{b} = 1 \emph{R} (red) and 0.6 \emph{R} (cyan)); (\emph{b}) the effects of airway pressure on the different between the maximum and the minimum of \emph{$P_{wall}$} and \emph{$\tau_{wall}$}.}
\label{fig:ka}
\end{figure}

The effect of thickness of liquid plug is also considered as shown in figure 11. In figure 11(\emph{a}), it can be seen that the aerosol mass is a linear increase as increase of the thickness of plug. Like the effect of the pressure drop, the thicker the liquid plug is, the larger the different between the maximum and the minimum of \emph{$P_{wall}$} and \emph{$\tau_{wall}$} are (see in figure 11(\emph{b})). This is because the rupture time delays when the plug thickens and the \emph{$P_{wall}$} and \emph{$\tau_{wall}$} are accumulated. But the increment of the different between the maximum and the minimum of \emph{$P_{wall}$} and \emph{$\tau_{wall}$} due to the plug thickness increases is smaller than that of due to the pressure drop increases. It means that the plug thickness has more effect on the aerosol mass than the pressure drop does, while the pressure drop has more effect on the cell injury than the plug thickness does.
\begin{figure}
  \centerline{\includegraphics{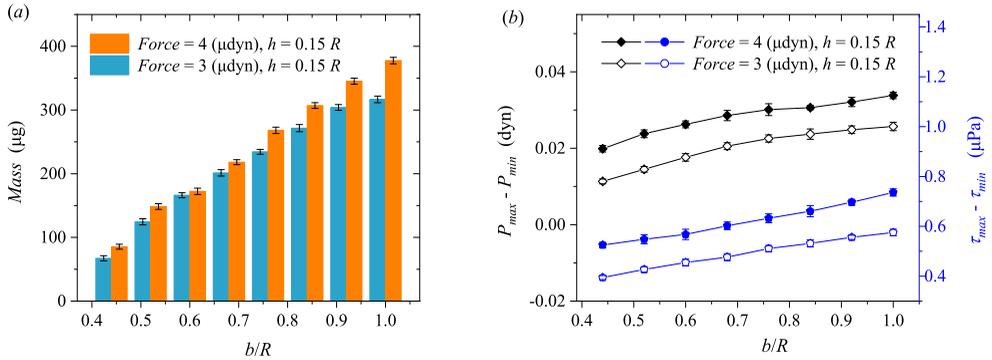}} 
  \caption{Effects of thickness of liquid plug \emph{b} on aerosol size (\emph{a}) the relationship between \emph{b} and aerosol quality (\emph{h} = 0.15 \emph{R}, \emph{Force} = 4 (orange) and 3 (blue) $ \mu $dyn); (\emph{b}) the effects of \emph{b} on the different between the maximum and the minimum of \emph{$P_{wall}$}; and the effects of \emph{b} on the different between the maximum and the minimum of \emph{$\tau_{wall}$}.}
\label{fig:ka}
\end{figure}
\begin{figure}
  \centerline{\includegraphics{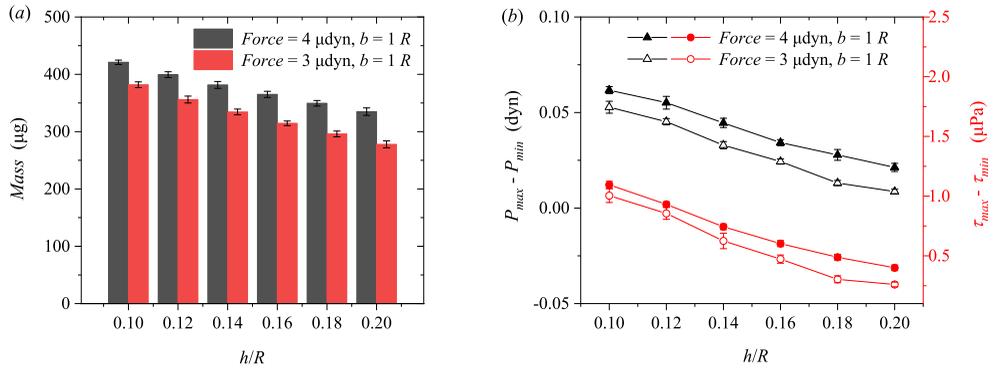}} 
  \caption{Effects of thickness of liquid film h on aerosol size (\emph{a}) the relationship between \emph{h} and aerosol quality (\emph{b} = 1 \emph{R}, \emph{Force} = 4 (dark grey) and 3 (red) $ \mu $dyn); (\emph{b}) the effects of \emph{h} on the different between the maximum and the minimum of \emph{$P_{wall}$}; and the effects of \emph{h} on the different between the maximum and the minimum of \emph{$\tau_{wall}$}.}
\label{fig:ka}
\end{figure}

The effect of thickness of liquid film is taken into account as shown in figure 12. Unlike the effects of the pressure drop and the plug thickness, the aerosol mass is a linear decrease as increase of the thickness of liquid film as shown in figure 12(\emph{a}). In figure 12(\emph{b}), it can been find that the different between the maximum and the minimum of \emph{$P_{wall}$} declines from near 0.06 dyn to near 0.02 dyn and that of \emph{$\tau_{wall}$} similarly reduces from near 1.0 $ \mu $Pa to near 0.25 $ \mu $Pa as the thickness of liquid film increases from 0.1 \emph{R} to 0.2 \emph{R} and the body force is 3 $ \mu $dyn. The results is agreement with the results of \cite{Muradoglu2019}. As the thickness of liquid film becomes thickens, the impact force of the broken liquid plug takes more time to reach the epithelium and weakens.

\section{Conclusion}\label{sec:types_paper}
The EOS-based multiphase flow model is used to numerical simulation the gas-liquid flow of pulmonary airway reopening. The numerical model is first validated with the simulations of \cite{Fujioka2008}. The result is in reasonable agreement with their result. Furthermore, two rupture cases with and without aerosol formation are contrasted and analyzed. It is found that the injury on the epithelium is resulted by the impact of the broken liquid plug by analyzing of the momentum distributions during liquid plug rupture. And the injury in the case with aerosol formation is essentially the same in the case without aerosol formation even while the pressure drop enlarges. It is because that a part of momentum is carried off by the aerosol after the plug ruptures. Then extensive simulations are performed to investigate the effects of pressure drop, thickness of liquid plug and film on aerosol size and the mechanical stresses. The results show that aerosol size and the mechanical stresses increase with the increase of pulmonary airway pressure as the thickness of the liquid plug and film keep constant. Similar, the aerosol size and the mechanical stresses increase with increasing the thickness of the liquid plug as pulmonary airway pressure and the thickness of liquid film keep constant. In contrast, the aerosol size and the mechanical stresses decrease with increasing the thickness of the liquid film as pulmonary airway pressure and the thickness of liquid plug keep constant.

This paper develops a new model capable of predicting the pulmonary airway reopening whit and without aerosol formation. The model will be further developed and used in observing the rupture under cyclic forcing and at airway bifurcation (\cite{Vaughan2016}; \cite{Mamba2018}).

\section*{Acknowledgments}
This work was supported by the National Natural Science Foundation of China (Grant No. 11862003, No. 81860635, and No. 12062005), the Key Project of Guangxi Natural Science Foundation (Grant No. 2017GXNSFDA198038), Guangxi “Bagui Scholar” Teams for Innovation and Research Project, and Guangxi Collaborative Innovation Center of Multi-source Information Integration and Intelligent Processing.


\bibliographystyle{jfm}
\bibliography{jfm}


\end{document}